\documentclass[twocolumn,pra,showpacs,amsmath,amssymb]{revtex4}

\usepackage{amsmath,amsfonts,amsthm,graphicx,color}

\def\be{\begin{equation}}
\def\ee{\end{equation}}
\def\bea{\begin{eqnarray}}
\def\eea{\end{eqnarray}}
\def\bma{\begin{mathletters}}
\def\ema{\end{mathletters}}

\def\0{\overline{0}}

\def\q0{\underline{0}}

\def\H{{\cal H}}

\DeclareMathOperator{\Ree}{Re}
\DeclareMathOperator{\Imm}{Im}

\def\tr{\mbox{tr}}
\def\one{\leavevmode\hbox{\small1\normalsize\kern-.33em1}}

\def\bra#1{\langle#1|} \def\ket#1{|#1\rangle}

\def\proj#1{\ket{#1}\!\bra{#1}}

\newcommand{\CC}{\mathbb{C}}
\newcommand{\RR}{\mathbb{R}}
\newcommand{\ZZ}{\mathbb{Z}}

\newcommand{\bk}[2]{\langle #1|#2\rangle}

\begin{document}

\title{ Grothendieck's constant and local models
for noisy entangled quantum states
 }

\author{Antonio Ac\'\i n$^{1}$, Nicolas Gisin$^2$ and
Benjamin Toner$^3$}

\affiliation{ $^1$ICFO-Institut de Ci\`encies Fot\`oniques,
Mediterranean Technology Park, 08860 Castelldefels (Barcelona), Spain\\
$^2$GAP-Optique, University of Geneva, 20, Rue de l'\'Ecole de
M\'edecine, CH-1211 Geneva 4, Switzerland\\
$^3$ Institute for Quantum Information, California Institute of
Technology, Pasadena, CA 91125, USA  }

\date{\today}


\begin{abstract}
  We relate the nonlocal properties of noisy entangled states to
  Grothendieck's constant, a mathematical constant appearing in Banach
  space theory.  For two-qubit Werner states
  $\rho^W_p=p\,\proj{\psi^-}+(1-p){\one}/{4}$, we show that there is a
  local model for projective measurements if and only if $p \le
  1/K_G(3)$, where $K_G(3)$ is Grothendieck's constant of order 3.
  Known bounds on $K_G(3)$ prove the existence of this model at least
  for $p \lesssim 0.66$, quite close to the current region of Bell
  violation, $p \sim 0.71$.  We generalize this result to arbitrary
  quantum states.
\end{abstract}


\maketitle

\section{Introduction}

The impossibility of reproducing all correlations observed in
composite quantum systems using models {\sl \` a la}
Einstein-Podolsky-Rosen (EPR)~\cite{EPR} was proven in 1964 by
Bell. In his seminal work~\cite{Bell}, Bell showed that all local
models satisfy some conditions, the so-called Bell inequalities,
but there are measurements on quantum states that violate a Bell inequality.
Therefore, we say that Quantum Mechanics is
nonlocal \cite{note}. Experimental verification
of Bell inequality violation closed the EPR debate, up to some
technical loopholes~\cite{exp}.

From an operational point of view it is not difficult to define
when a quantum state exhibits nonclassical correlations.  Suppose
that two parties, Alice (A) and Bob (B), share a mixed quantum
state $\rho$ with support on $\H_A\otimes\H_B$, where $\H_A$
($\H_B$) is the local Hilbert space of A's (B's) system.  Then
$\rho$ contains quantum correlations when its preparation requires
a nonlocal quantum resource. Conversely, a quantum state is
classically correlated, or separable, when it can be prepared
using only local quantum operations and classical communication
(LOCC). From this definition, due to Werner \cite{werner}, it
follows that a
quantum state $\rho$ is separable if it can be expressed as a
mixture of product states,
$\rho=\sum_{i=1}^N p_i\proj{\psi_A^i}\otimes\proj{\psi_B^i}$.
A state that cannot be written in this form has quantum
correlations and is termed entangled.
But the above definition, in spite of its clear physical meaning,
is somewhat impractical.  Tests to distinguish separable from
entangled states are complicated~\cite{Doherty}, except when
$d_A=2$ and $d_B\leq 3$ \cite{PPT}, $d_A$ and $d_B$ denoting the
dimensions of the local subsystems.

Violation of a Bell inequality by a quantum state is, in many
situations, a witness of useful correlations~\cite{useful}.  In
particular, Bell inequality violation is a witness of a quantum
state's entanglement. Now, the question is: Are all entangled states
nonlocal? For the case of pure states, the answer is yes~\cite{gisin}:
all entangled pure states violate the CHSH inequality \cite{CHSH}.  In
1989, Werner showed that the previous result cannot be generalized to
mixed states~\cite{Wernernote}.
He introduced what are now called Werner states, and
gave a local hidden variables (LHV) model for measurement outcomes for
some entangled states in this family \cite{werner}. Although the
construction only worked for projective
measurements, his result has since been
extended to general measurements~\cite{barrett}.

\begin{figure}
  \includegraphics[width=8cm]{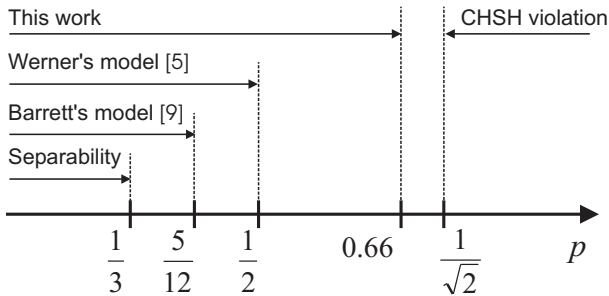}\\
  \caption{Non-local properties of two-qubit Werner states, $\rho^W_p$.
  Werner's local model works up to $p=1/2$, while the CHSH inequality is violated
  when $p>2^{-1/2}\sim 0.71$. Here, we prove the existence
  of a local model for projective measurements when $p\lesssim 0.66$.}\label{summary}
\end{figure}

In spite of these partial results, it is in general extremely
difficult to determine whether an entangled state has a local model
or not~\cite{altdef}, since (i) finding all Bell
inequalities is a computationally hard problem \cite{Pitowski,AlonNaor} and
(ii) the number of possible measurement is unbounded (see however
\cite{TDS} for recent progress). This question remains unanswered even in the simplest
case of Werner states of two qubits. These are mixtures of the
singlet $\ket{\psi^-}=(\ket{01}-\ket{10})/\sqrt 2$ with white
noise of the form
\begin{equation}\label{wstates}
    \rho^W_p=p\,\proj{\psi^-}+(1-p\,)\frac{\one}{4}.
\end{equation}
It is known that Werner states are separable iff $p \le 1/3$,
admit a LHV model for all measurements for $p \le
5/12$~\cite{barrett}, admit a LHV for projective measurements for
$p \le 1/2$~\cite{werner}
and violate the CHSH inequality for $p>1/\sqrt 2$
(see Fig.~\ref{summary}). However, the critical value of $p$,
denoted $p_c^W$, at which two-qubit Werner states cease to be
nonlocal under projective measurements is unknown. This question
is particularly relevant from an experimental point of view, since
$p_c^W$ specifies the amount of noise the singlet tolerates
before losing its nonlocal properties.

In this paper, we exploit the connection between correlation Bell
inequalities and Grothendieck's constant~\cite{finch}, first
noticed by Tsirelson~\cite{cirelson}, to prove the existence of a
local model for several noisy entangled states. We first
demonstrate that $p_c^W$ is related to a generalization of this
constant, namely, $p_c^W = 1/K_G(3)$, where $K_G(3)$ is
Grothendieck's constant of order 3~\cite{krivine}. The exact value of
$K_G(3)$ is unknown, but known bounds establish that
$0.6595 \le p_c^W \le 1/\sqrt 2$.  Thus, we close more
than three-quarters of the gap between Werner's result and the
known region of Bell inequality violation (see Fig.
\ref{summary}). 
Next, we show that if Alice (or Bob) is restricted to make
measurements in a plane of the Poincar\'e sphere, then there is an
explicit LHV model for all $p\le 1/K_G(2) = 1/\sqrt 2$.  This
improves on the bound of Larsson, who constructed a LHV model for
planar measurements for $p \leq 2/\pi$~\cite{Larsson}.  Thus, in
the case of planar projective measurements, violation of the CHSH
inequality completely characterizes the nonlocality of two-qubit
Werner states.

In the case of {\em traceless two-outcome} observables, we can
extend our results to mixtures of an arbitrary state $\rho$ on
$\CC^d \otimes \CC^d$ with the identity, of the form \cite{noted}
\begin{equation}
\label{genst}
    \rho_p=p\, \rho + (1-p\,)\frac{\one}{d^2}.
\end{equation}
Denote by $p_c(\rho)$ the maximum value of $p$ for which there
exists a LHV model for the joint correlation of traceless
two-outcome observables on $\rho_p$, and define
\begin{equation}
\label{critpr}
  p_c^d = \min_\rho p_c(\rho) \quad\quad
  p_c =  \lim_{d\to \infty}p_c^d .
\end{equation}
Then $p_c = 1/K_G$ where $K_G$ is Grothendieck's constant. Again,
the exact value of $K_G$ is unknown, but known
bounds imply $0.5611\le p_c\le 0.5963$.  

Finally, we discuss the opposite question of finding Bell
inequalities better than the CHSH inequality at detecting the
nonlocality of $\rho_p^W$, or, more generally, of Bell diagonal
states~\cite{note2}. In particular, we show that none of the
$I_{nn22}$ Bell inequalities introduced in Ref. \cite{CG} is
better than the CHSH inequality for these states.



Before proving our results, we require some notation. We write a
two-outcome measurement by Alice (resp. Bob) as $\{{A}^+, {A}^-\}$
(resp. ${\{{B}^+, {B}^-\}}$), where the projectors ${A}^\pm$
correspond to measurement outcomes $\pm 1$. We define the {\em
observable} corresponding to Alice's (Bob's) measurement as ${A} =
{A}^+ - {A}^-$ (${B} = {B}^+ - {B}^-$).  An observable $A$ is {\em
traceless} if $\tr A = 0$, or equivalently $\tr A^- = \tr A^+$.
The {\em joint correlation} of Alice and Bob's measurement
results, denoted $\alpha$ and $\beta$ respectively, is
\begin{equation}
    \label{eq:correl}
    \langle \alpha\beta \rangle = \tr\left( {A} \otimes {B}
    \, {\rho} \right).
\end{equation}
Alice's {\em local marginal} is specified by $\langle \alpha
\rangle = \tr\left( {A} \otimes {\one} \, {\rho} \right)$, and
similarly for Bob. Together, $\langle \alpha\beta \rangle$,
$\langle \alpha\rangle$ and $\langle \beta \rangle$ define the
full probability distribution for two-outcome measurements on $\bf
\rho$.  A LHV model for the full probability distribution is one
that gives the same values $\langle \alpha\beta \rangle$, $\langle
\alpha\rangle$ and $\langle \beta \rangle$ as quantum theory.  A
LHV model for the joint correlation is one that gives the same
joint correlation $\langle \alpha \beta \rangle$, but not
necessarily the correct marginals. In the qubit case, the
projective measurements applied by the parties are specified by
the direction of their Stern-Gerlach apparatuses, given by
normalized three-dimensional real vectors $\vec a$ and $\vec b$:
${A} = \vec a \cdot \vec {\sigma} $ and ${B} = \vec b \cdot \vec
{\sigma}$.

\section{Werner states}

Let us first consider the case of Werner states (\ref{wstates}).
For projective measurements on $\rho^W_p$, LHV simulation of the
joint correlation is sufficient to reproduce the full probability
distribution.  This follows from:

{\bf Lemma~1}: Suppose that there is a LHV model $L$ that gives
joint correlation $\langle \alpha \beta \rangle_{{L}}$. Then there is a
LHV model $L'$ with the same joint correlation and uniform marginals:
$\langle \alpha \beta \rangle_{{L'}} = \langle \alpha \beta
\rangle_{{L}}$, $\langle \alpha \rangle_{L'} = \langle \beta
\rangle_{L'} = 0$.

{\it Proof}: Let $\alpha$ and $\beta$ be the outputs generated by
the LHV $L$ (dependent on the hidden variables and measurement
choices).  Define a new LHV $L'$ by augmenting the hidden variables of
$L$ with an additional random bit $c \in \{-1, 1\}$.  In $L'$, Alice outputs $c
\alpha$ and Bob $c \beta$.  $\Box$

Therefore, the analysis of the non-local properties of Werner
states under projective measurements can be restricted to Bell
inequalities involving only the joint correlation.  Actually, this
holds for any Bell diagonal state, under projective measurements,
since $\tr_A\rho=\tr_B\rho=\one/2$ for all these states, so all
projective measurements give uniform marginals. In the Bell
scenarios we consider, Alice and Bob each choose from $m$
observables, specified by $\{A_1,\ldots,A_m\}$ and
$\{B_1,\ldots,B_m\}$. We can write a generic correlation Bell
inequality as
\begin{equation}\label{corrBell}
    |\sum_{i,j=1}^m M_{ij}\, \langle \alpha_i \beta_j \rangle|\leq 1 ,
\end{equation}
where $M = (M_{ij})$ is a $m\times m$ matrix of real coefficients
defining the Bell inequality.  The matrix $M$ is normalized such that
the local bound is achieved by a deterministic local model, i.e.,
\begin{equation}
  \max_{a_i=\pm 1,\,b_j=\pm 1}|\sum_{i,j=1}^m M_{ij}\,a_ib_j| = 1.
\end{equation}
For the singlet state $\langle \alpha_i \beta_j\rangle_{\Psi^-}=-\vec
a_i\cdot \vec b_j$.  We obtain the maximum ratio of Bell inequality
violation for the singlet state, denoted $Q$, by maximizing over
normalized Bell inequalities, and taking the limit as the number of
settings goes to infinity:
\begin{equation}\label{singlviol}
    Q =\lim_{m \to \infty}\sup_{M_{ij}}\
    \max_{\vec a_i,\,\vec b_j} |\sum_{i,j=1}^m M_{ij}\,\vec a_i\cdot\vec
    b_j|.
\end{equation}
Since all joint correlations vanish for the maximally-mixed state,
it follows that the critical point at which two-qubit Werner states do not
violate any Bell inequality is $p_c^W=1/Q$.

As first noticed by Tsirelson, the previous formulation of the Bell
inequality problem is closely related to the definition of
Grothendieck's inequality and Grothendieck's constant, $K_G$ (see
\cite{cirelson} for details).  Grothendieck's inequality first arose
in Banach space theory, particularly in the theory of
$p$-summing operators~\cite{groth}.  We shall need a refinement of his
constant, which can be defined as follows~\cite{finch}:

{\bf Definition 1:} 
For any integer $n \geq 2$, Grothendieck's
constant of order $n$, denoted $K_G(n)$, is the smallest number with the following
property: Let $M$ be any $m \times m$ matrix for which
\begin{equation}\label{groth1}
    |\sum_{i,j=1}^m M_{ij}\,a_ib_j|\leq 1,
\end{equation}
for all real numbers $a_1,\ldots,a_m,b_1,\ldots,b_m \in [-1,+1]$.
Then
\begin{equation}\label{groth2}
|\sum_{i,j=1}^m M_{ij}\,\vec a_i\cdot\vec b_j|\leq K_G(n) ,
\end{equation}
for all unit vectors $\vec a_1,\ldots,\vec a_m,\vec b_1,\ldots,\vec b_m$ in $\RR^n$.



{\bf Definition 2:} Grothendieck's constant is defined as
\begin{equation}
K_G = \lim_{n \to \infty} K_G(n).
\end{equation}

The best bounds currently known for $K_G$ are $1.6770\le K_G\le
\pi/(2\log(1+\sqrt 2))=1.7822$~\cite{FR}. The lower bound is due
to Reeds and, independently, Davies~\cite{RD}, while the upper
bound is due
to Krivine~\cite{krivine}.

It follows immediately from the first definition that the maximal
Bell violation for the singlet state (\ref{singlviol}) is
$K_G(3)$. We have therefore proved

{\bf Theorem 1:} \label{theorem:result} There is a LHV model for
projective measurements on the Werner state $\rho_p^W$ if and only
if $p \le p_c^W={1}/{K_G(3)}$.

It is known that $\sqrt 2 \le K_G(3) \le 1.5163$.  The lower bound
follows from the CHSH inequality; the upper bound is again due to
Krivine~\cite{krivine}.  He shows that $K_G(3)\leq\pi/(2c_3)$
where $c_3$ is the unique solution of
\begin{equation}
\label{groth3}
  \frac{\sqrt c_3}{2}\, \int_0^{c_3} t^{-3/2}\sin t\, dt = 1
\end{equation}
in the interval $[0,\pi/2]$.  Numerically we find that $c_3
\approx 1.0360$.  This implies $K_G(3) \leq 1.5163$ and
$p_c^W\geq 0.6595$.
Furthermore,  it turns out that an explicit LHV model emerges from
Krivine's upper bound on $K_G(3)$, and the details are presented
in~\cite{Toner1}.

Another result
 follows from Krivine's work:

{\bf Theorem 2:} If Alice's projective measurements are restricted
to a plane in the Poincar\'e sphere, then there is a LHV model for
$\rho_p^W$ if and only if $p \le 1/\sqrt 2$.

{\sl Proof:} In this case, the vectors $\vec a_i$ in
(\ref{singlviol}) are two-dimensional.  Since the quantum
correlation 
depends only on the projection of $\vec b_j$ onto $\vec a_i$, we
can assume that the vectors $\vec b_j$ lie in the same plane.
%
It follows that $p_c^W= 1/K_G(2)$ for planar measurements, and
Krivine has shown that $K_G(2)$ is equal to $\sqrt 2$~\cite{krivine}. $\Box$

Again Krivine's proof can be adapted to give an explicit LHV model
for planar measurements, valid for $p \leq 1/\sqrt
2$~\cite{Toner1}.

\section{Generalization to higher dimension}

It is possible to extend these results to general states of the
form (\ref{genst}), if we restrict our analysis to correlation
Bell inequalities of {\it traceless} two-outcome observables.
Admittedly, this analysis is far from sufficient. Indeed it
does not allow us to determine whether the full probability
distribution admits a LHV model even in the case of two-outcome
measurements, since the most general Bell inequalities have terms
that depend on marginal probabilities \cite{CG}. Mindful of this
caveat, we now prove the existence of LHV models for the
joint correlation of the states (\ref{genst}). To make
the connection with Grothendieck's constant, we start with a
representation of quantum correlations as dot products, first
noted by Tsirelson~\cite{cirelson}.
It is sufficient to restrict to the case of pure states, since we can
obtain a LHV for a mixed state $\rho$ by decomposing it into a convex
sum of pure states, and taking a convex combination of the LHV's for
those pure states.

{\bf Lemma~2}: Suppose Alice and Bob measure observables $A$ and
$B$ on a pure quantum state $\ket \psi \in \CC^d \otimes \CC^d$.
Then we can associate a real unit vector $\vec a\in \RR^{2d^2}$
with $A$ (independent of ${B}$), and a real unit vector $\vec b
\in \RR^{2d^2}$ with $B$ (independent of $A$) such that $  \langle
\alpha \beta \rangle_{\psi } = \vec a \cdot \vec b$. Moreover, if
$\ket \psi$ is maximally entangled, then we can assume the vectors
$\vec a$ and $\vec b$ lie in $\RR^{d^2-1}$.

{\it Proof}: Let $\ket a = A \otimes \one_B \ket \psi$ and $\ket b = \one_A
  \otimes B \ket \psi$.  Then $\langle \alpha \beta \rangle =
  \bk ab$, $\bk aa = \bk bb = 1$.  Denote the components of $\ket a$
  as $a_i$ where $i = 1, 2,\ldots,d^2$, and similarly for $\ket b$.
  We now define a $2d^2$--dimensional real vector $\vec a =$ $(\Ree a_1$,
  $\Imm a_1$, $\Ree a_2$, $\Imm a_2$, \dots, $\Ree a_{d^2}$, $\Imm a_{d^2})$,
  and similarly $\vec b =$ $(\Ree b_1$, $\Imm b_1$, $\Ree b_2$, $\Imm b_2$,
  \dots, $\Ree b_{d^2}$, $\Imm b_{d^2})$.  Then $\vec a \cdot \vec a =
  \vec b \cdot \vec b = 1$ and $\langle \alpha \beta \rangle = \vec a
  \cdot \vec b$ (because $\bk ab$ is real).

If $\ket \psi$ is maximally entangled, we can assume $\ket \psi = \ket
{\psi^+} = 1/\sqrt d\,\sum_{i=1}^d \ket {ii}$. We calculate
$\langle \alpha \beta \rangle_{\psi^+} = \tr_A \left({A}{B}^t \right)/d$
where ${B}^t$ is the transpose of $B$.  Introduce a
$(d^2-1)$--dimensional basis
${g}_i$ for traceless operators on ${\cal H}_A$, normalized such
that $\tr \left( {g}_i {g}_j \right) = d \delta_{ij}$.  Let ${A} =
\sum_i a_i {g}_i$, ${B}^t = \sum_i b_i {g}_i$, which define the
vectors $\vec a$ and $\vec b$.  Squaring these
definitions and taking the trace gives $\sum_i a_i^2 = \sum_i b_i^2 =
1$. Finally, $\tr \left({A}{B}^t \right) = d
\sum_i a_i b_j$, which implies that $\langle \alpha \beta \rangle =
\sum_i a_i b_i = \vec a \cdot \vec b$.

The converse of Lemma~2 is also true: all dot products of
normalized vectors, $\vec a,\vec b\in\RR^{n}$, are realized as
observables on $\ket{\psi^+}$, where $n=2 \lfloor \log_2 d \rfloor
+ 1$. This result was derived by Tsirelson in
Ref.~\cite{cirelson}. For the sake of completeness, we state it
here without proof (see \cite{cirelson} for the details).

 {\bf Theorem 3 \cite{cirelson}:}
 Let $\{\hat a_i\}_{i=1}^m$
   and $\{\hat b_j\}_{j=1}^m$ be sets of
   unit vectors in $\RR^n$.  Let $d = 2^{\lfloor n/2 \rfloor}$ and
   $\ket \Phi$ be a maximally entangled state on $\CC^d \otimes
   \CC^d$.  Then there are observables $A_1\ldots,
   A_m$ and $B_1\ldots,B_m$ on $\CC^d$ such that
 \begin{eqnarray}
 \langle \alpha_i \rangle &=& \bra \Phi {A}_i\otimes \one \ket \Phi = 0,\\
 \langle \beta_j \rangle &=& \bra \Phi \one \otimes {B}_j \ket \Phi = 0,\\
 \langle \alpha_i \beta_j \rangle &=& \bra \Phi {A}_i \otimes {B}_j
 \ket \Phi = \hat a_i \cdot \hat b_j,
 \end{eqnarray}
 for all $1 \leq i,j \leq m$.

Note that in our case, the stipulation that the observables be
traceless ensures that their outcomes are random on the maximally
mixed state.  It follows from Lemma~2 and Theorem~3 that

{\bf Theorem 4:} Let $\rho$ be a state on $\CC^d \otimes \CC^d$
and define $\rho_p$ and $p_c^d$ as in Eqs.
(\ref{genst},\ref{critpr}). Then
\begin{equation}
\frac{1}{K_G(2 d^2)}\leq p_c^d \leq \frac{1}{K_G(2
  \lfloor \log_2 d \rfloor + 1)}.
\end{equation}
In other words, there is always a LHV model for the joint correlation of
traceless two-outcome observables on $\rho_p$ for $p \leq 1/K_G(2
d^2)$ and there is a state (in fact, the maximally entangled state on
$\lfloor \log_2 d \rfloor$ qubits) such that the
joint correlation is nonlocal for $p > 1/K_G(2 \lfloor \log_2 d
\rfloor + 1)$.


{\bf Corollary 1:} The threshold noise for the joint correlation of
two-outcome traceless observables is $p_c = 1/K_G$.

This follows from the previous theorem, taking the limit $d \to
\infty$. The known bounds imply $0.5611\le p_c \le 0.5963$.
Compare this to $p_s$, the threshold noise at which the state
$\rho_p$ is guaranteed separable: while $p_s$ decreases with dimension
at least as $1/(1+d)$ \cite{GB}, $p_c$ approaches a constant.  
In the case of two-qubit systems, we can be more specific,
because projective measurements are
traceless and have two outcomes:

{\bf Corollary 2:} Suppose $\rho$ is an arbitrary state on $\CC^2 \otimes
\CC^2$.  Then there is a LHV model for the joint correlation on
$\rho_p =  p\, \rho +(1-p){\one}/{4}$ for $p\le 1/K_G(8)$. In particular, $K_G(8)\le 1.6641$
\cite{krivine,Toner1}, which implies there is a LHV model for
$p\le0.6009$.

For maximally entangled states, marginals of traceless
observables are uniform, so Lemmas~1 and 2 imply:

{\bf Theorem 5}:  Let $\rho_p = p\, \proj {\psi^+} +
(1-p)\one/d^2$ where $\ket {\psi^+}$ is a maximally entangled
state in $\CC^d \otimes \CC^d$.  Then there is a LHV for the full
probability distribution arising from traceless observables for $p
\leq 1/{K_G(d^2-1)}$.


\section{Bell inequalities for Werner states}

Just as upper bounds on $K_G(n)$ yield LHV models, lower bounds
yield Bell inequalities.  The case of Werner states appears of
particular interest: at present, there is no Bell inequality
better than CHSH at detecting the nonlocality of $\rho_p^W$
\cite{open}. This and other approaches to construct new Bell
inequalities will be presented in \cite{Toner1}. Unfortunately,
none of these inequalities could be proven to be better than CHSH.
It is remarkable how difficult it is to enlarge this region of
Bell violation or, equivalently, to show that $K_G(3)>K_G(2)=\sqrt
2$. Actually, in the case of random marginal probabilities, as for
Bell diagonal states under projective measurements, no improvement
over the CHSH inequality can be obtained using $3\times n$
measurements \cite{garg}.

A similar result can also be proven for the whole family of the
so-called $I_{nn22}$ \cite{CG} Bell inequalities. These are
specified by a matrix of zeros and $\pm 1$ as follows,
\begin{equation}\label{inn22}
I_{nn22} = \left(
\begin{tabular}{c||cccccc}
  & -1 & 0 & $\cdots$ & $\cdots$ & $\cdots$ & 0\\
\hline \hline
 -(n-1) & 1 & $\cdots$ & $\cdots$ & $\cdots$ & $\cdots$ & 1\\
 -(n-2) & 1 & $\cdots$ & $\cdots$ & $\cdots$ & 1 & -1\\
 -(n-3) & 1 & $\cdots$ & $\cdots$ & 1 & -1 & 0\\
 $\vdots$ & $\vdots$ & $\vdots$ & $\vdots$ & $\vdots$ & $\vdots$ & $\vdots$\\
 -1 & 1 & 1 & -1 & 0 & $\cdots$ & 0\\
 0 & 1 & -1 & 0 & $\cdots$ & $\cdots$ & 0\\
\end{tabular}
\right).
\end{equation}
All the coefficients in the first column (row) refer to Alice's
(Bob's) marginal probabilities, while the rest of terms are for
joint probabilities. Only one of the two possible outcomes, say
$+1$, appears in the inequality and its local bound is always
zero. For example, when $n=2$, and denoting
$p\,(a_i,b_j)=p\,(a_i=+1,b_j=+1)$, $I_{2222}$ reads
\begin{eqnarray}\label{i2222}
    &&p\,(a_1,b_1)+p\,(a_1,b_2)+p\,(a_2,b_1)-p\,(a_2,b_2)- \nonumber\\
    &&p\,(a_1=+1)-p\,(b_1=+1)\leq 0 ,
\end{eqnarray}
which is equivalent to the CHSH inequality.

{\bf Theorem 6:} Consider the set of $I_{nn22}$ Bell inequalities,
for $n$ two-outcome settings. Then, if a Bell diagonal state
violates any of these inequalities with projective measurements,
it also violates the CHSH inequality.

{\sl Proof:} Our proof takes advantage of the fact that all
marginal probabilities for projective measurements on Bell
diagonal states are fully random. Thus, when dealing with these
states, one can put all the terms in the first row and column of
(\ref{inn22}) equal to $1/2$ 
. In order to avoid confusion, we denote by $I'_n$ the $I_{nn22}$
inequalities where the local terms have been replaced by 1/2.

We start our proof with the simplest non-trivial case $I_{3322}$.
For Bell diagonal states, it can be written as
\begin{equation}
    I'_3=\frac{1}{2}\left(I'_2(1213)+I'_2(1223)
    +I'_2(1312)+I'_2(2312)\right)\leq 0 ,
\end{equation}
where the arguments of $I'_2(ijkl)$ are the measurements that
appear in the $I'_2$ inequality, $i$ and $j$ for Alice, and $k$ and
$l$ for Bob.  From this identity we have that the violation of $I'_3$
implies that at least one of the $I'_2$ inequalities is violated too.
This procedure can be generalized for all $n$: the idea is to express
$I'_n$ in terms of $I'_2$ inequalities using the joint probability
terms with a negative sign in (\ref{inn22}). For example, when $n=4$
one has
\begin{eqnarray}\label{i4bd}
    I'_4&=&\frac{1}{3}\left[I'_2(1214)+I'_2(1224)+I'_2(1234)+
    I'_2(1313)+\right.\nonumber\\
    &&\left.I'_2(1323)+I'_2(2313)+I'_2(2323)+
    I'_2(1412)+\right.\nonumber\\
    &&\left.I'_2(2412)+I'_2(3412)+p\,(a_3,b_3)-\frac{1}{2}\right]\leq
    0 .
\end{eqnarray}
Note that since all local probabilities are equal to $1/2$,
$p\,(a_3,b_3)-1/2$ is never positive. Thus, whenever $I'_4>0$, at
least one of the $I'_2$ inequalities appearing in (\ref{i4bd}) is
violated. For arbitrary $n$, $I'_n$ can always be written as
\begin{equation}\label{inbd}
    I'_n=\frac{1}{n-1}\left[\sum_{i=1}^{s_1(n)} I'_2+
    \sum_{i=1}^{s_2(n)}\left(p\,(a,b)-\frac{1}{2}\right)\right]\leq 0 ,
\end{equation}
i.e. the sum of $s_1(n)$ $I'_2$ inequalities and $s_2(n)$ negative
terms $p\,(a_i,b_j)-1/2$, up to an $n-1$ factor. Some patient
calculation shows that $s_1(n)=n(n^2-1)/6$ and
$s_2(n)=(n-1)(n-2)(n-3)/6$. Thus, if a Bell diagonal state
violates $I_{nn22}$, it also violates a CHSH inequality.
Consequently, none of these inequalities enlarge the known region
of Bell violation for Werner states.

After seeing these results, one would be tempted to conjecture
that the CHSH violation provides a necessary and sufficient
condition for detecting the nonlocality of Bell diagonal states,
and in particular of Werner states. This result, however, would
imply that $K_G(3)=K_G(2)=\sqrt 2$, which seems unlikely.
Actually, one can find in \cite{FR} an explicit construction with
20 settings showing that $K_G(5)\geq 10/7>\sqrt 2$. More recently,
one of us has shown that $K_G(4)>\sqrt 2$ as well~\cite{Toner1}.

\section{Conclusions}

In this work, we have exploited the connection between Bell
correlation inequalities and Grothendieck's constants to prove the
existence of LHV models for several noisy entangled states. In the
case of Werner states, one can demonstrate the existence of a
local model for projective measurements up to $p\sim 0.66$, close
to the known region of Bell violation. Although we only proved
here the existence of the LHV models, the correspondence between
noise thresholds and Grothendieck's constants can also be
exploited to construct the {\it explicit} models. Indeed, these
can be extracted from (the proofs of) Krivine's upper bounds on
$K_G(n)$. The details are presented in Ref.~\cite{Toner1}.

\section{Acknowledgements}

This work is supported by the National Science Foundation under
grant EIA-0086038, a Spanish MCyT ``Ram\'on y Cajal" grant, the
Generalitat de Catalunya, the Swiss NCCR ``Quantum Photonics" and
OFES within the European project RESQ (IST-2001-37559).  We thank Steven Finch for
providing us with Ref.~\cite{RD}.


\begin{references}
\bibitem{note}
By nonlocality we refer to the impossibility that quantum
correlations can be described by a local model. 
\bibitem{noted}
By extending the smaller of $\H_A$ and $\H_B$, we can assume that
the local spaces have the same dimension $d = \max(d_A, d_B)$.
\bibitem{note2}
A state is Bell diagonal when its eigenvectors define a Bell
basis, $\ket{\phi^\pm}=(\ket{00}\pm\ket{11})/\sqrt 2$ and
$\ket{\psi^\pm}=(\ket{01}\pm\ket{10})/\sqrt 2$. Werner states are
Bell diagonal.
\end{references}
\end{document}